\documentclass[12pt,a4paper]{article}
\pdfoutput=1
\usepackage{jheppub}
\usepackage{amsfonts}
\usepackage{amscd}
\usepackage{amssymb}
\usepackage{amsmath}
\usepackage{epsfig}
\usepackage{latexsym}
\usepackage{mathtools}
\usepackage{hyperref}
\usepackage[vcentermath]{youngtab}
\hypersetup{
    colorlinks=true,
    linkcolor=black,
    citecolor=black,
    filecolor=black,
    urlcolor=black,}

\def \be  {\begin{equation}}
\def \ee  {\end{equation}}
\def \ba  {\begin{eqnarray}}
\def \ea  {\end{eqnarray}}
\def \bb  {\begin{thebibliography}}
\def \eb  {\end{thebibliography}}
\def \lab #1 {\label{#1}}

\newcommand\cI{\mathcal{I}}

\newcommand\cN{\mathcal{N}}
\newcommand\cO{\mathcal{O}}

\newcommand\cQ{\mathcal{Q}}
\newcommand\cR{\mathcal{R}}
\newcommand\cS{\mathcal{S}}
\newcommand\cT{\mathcal{T}}

\newcommand\cY{\mathcal{Y}}

\newcommand\SU{\mathrm{SU}}

\newcommand\rd{\mathrm{d}}
\newcommand\e{\mathrm{e}}

\newcommand\lb{\lambda}

\newcommand\la{\langle}
\newcommand\ra{\rangle}

\newcommand\tr{\mathrm{Tr}}

\newcommand\ir{\mathrm{IR}}
\newcommand\uv{\mathrm{UV}}

\title{Surface defects, the superconformal index and q-deformed Yang-Mills}
\author[a]{Luis F. Alday,}
\author[a]{Mathew Bullimore,}
\author[a]{Martin Fluder,}
\author[a,b]{Lotte Hollands}
\affiliation[a]{Mathematical Institute, University of Oxford,\\ 24-29 St Giles', Oxford, OX1 3LB, UK.}
\affiliation[b]{California Institute of Technology, Pasadena, CA 91125, USA.}

\abstract{ Recently a prescription to compute the four-dimensional ${\cal N}=2$ superconformal index in the presence of certain BPS surface defects has been given. These surface defects are labelled by symmetric representations of $SU(N)$. In the present paper we give a prescription to compute the superconformal index in the presence of surface defects labelled by arbitrary representations of $SU(N)$. Furthermore, we extend the dictionary between the ${\cal N}=2$ superconformal Schur-index and correlators of 
q-deformed Yang-Mills to incorporate such surface defects.
}

\begin{document}
\maketitle

\section{Introduction}

In \cite{Gaiotto:2009we} a large family of four-dimensional ${\cal N}=2$ supersymmetric gauge theories was introduced. These theories arise from compactifying the six-dimensional $(2,0)$ superconformal theory of type $A_{N-1}$ on a Riemann surface $C$ with punctures. This hints at a relation between supersymmetric observables in the 4d theory and quantities computed on the Riemann surface. An example of such a correspondence is the equivalence between the partition function of the 4d theories on $S^4$ and a Liouville/Toda correlator on the Riemann surface \cite{Alday:2009aq}. Another example, relevant for the present paper, is the equivalence between the superconformal index, or partition function on $S^1 \times S^3$,  and the correlator of a 2d topological QFT (TQFT) on the Riemann surface \cite{Gadde:2009kb}. The superconformal index depends on three superconformal fugacities (or simply fugacities) $(p,q,t)$. In a particular slice $(0,q,q)$, the TQFT was shown to be given by q-deformed 2d YM in the zero area limit \cite{Gadde:2011ik}. 

Four-dimensional ${\cal N}=2$ theories can also be decorated with supersymmetric defects. A natural question is how to compute the above 4d observables in the presence of these defects and to understand what they correspond to from the 2d perspective. These defects descend from defects on the parent 6d theory. For this paper, the relevant defects are codimension-four and located at a point in the Riemann surface, which produce a surface defect in the 4d theory. The partition function of 4d ${\cal N}=2$ theories in the presence of the corresponding surface operators was considered in \cite{Alday:2009fs}. From the 2d perspective, this is equivalent to considering the fusion of a special puncture (a degenerate Liouville primary) with the usual punctures. 

An elegant prescription to compute the index in the presence of certain surface defects was given in \cite{Gaiotto:2012xa}: adding a surface defect to a given theory ${\cal T}_{IR}$ amounts to acting on its index with a certain difference operator. This difference operator acts by shifting the flavor fugacities of one of the $SU(N)$ flavor punctures. Generalized S-duality implies that one should get the same result independently of which puncture the difference operator acts on. This has the powerful consequence that the index is diagonalized by eigenfunctions of those operators. This has led to a full determination of the index of 4d ${\cal N}=2$ theories of type $A_{N-1}$ and to a proof of the relation to 2d TFT mentioned above. 

The difference operators considered in \cite{Gaiotto:2012xa} are labelled by symmetric representations of $SU(N)$. They are self-adjoint with respect to the appropriate measure and they commute with each other. Furthermore, for the case of $A_1$ 
they have been shown to be closely related to the Hamiltonians of the elliptic Ruijsenaars-Schneider model \cite{Ruijsenaars:1986vq}. In this paper we explore the properties of these operators for higher rank groups. Closure under composition leads us to consider more general difference operators, labelled by arbitrary representations of $SU(N)$, which we construct explicitly. Applying these difference operators should correspond to computing the index in the presence of certain defect operators labelled by arbitrary irreducible representations. The difference operators obey simple composition rules and show some additional structure. For instance, each of them is given as a sum of terms, where each term can be associated to a semi-standard Young tableau. 

In the second part of the paper we turn our attention to the relation between the index and 2d q-YM and interpret the difference operators above as corresponding to the insertion of gauge invariant operators constructed from the scalar field of 2d q-YM, making precise the proposal of  \cite{Gaiotto:2012xa}.

The organization of this paper is as follows. In the next section we review the construction of \cite{Gaiotto:2012xa}, extending some of their results to higher rank. In section three we consider the composition of the difference operators considered in \cite{Gaiotto:2012xa}, in the so called Schur limit (a particular choice of the superconformal fugacities). By an appropriate similarity transformation, these composition rules become extremely simple, and the difference operators are constructed explicitly. In section four the complete dictionary between these difference operators and objects in 2d q-YM is built. The body of the paper ends with some conclusions and open problems. Appendix A contains a review of the group theory relevant for our computations while in appendix B we give the precise relation between our surface operators and the so-called q-difference operators considered in the mathematics literature.

\section{Superconformal index and surface defects}
\label{section:indexdefects}

In this paper we consider the index of superconformal $\cN=2$ theories in four dimensions. The superconformal index is a trace over the states of a superconformal field theory in radial quantization, that is, a twisted partition function on $S^1\times S^3$~\cite{Kinney:2005ej}. Furthermore, we concentrate on a particular limit of the superconformal index considered in~\cite{Gadde:2011ik,Gadde:2011uv}. This is defined by the trace
\be
\cI  = \tr (-1)^F q^{j_2-j_1+R}\prod_j a_j^{f_j}
\label{schurindex}
\ee 
over states in theory in radial quantization that are annihilated by two supercharges $\bar\cQ_{1\dot{-}}$ and $\cQ_{1+}$ and their hermitian conjugates
\be
\begin{aligned}
\left\{ \bar\cQ_{1\dot{-}}, (\bar\cQ_{1\dot{-}})^{\dagger} \right\} & = E-2j_2-2R+r \\
\left\{ \cQ_{1+},(\cQ_{1+})^\dagger \right\} & = E+2j_1-2R-r \, ,
\end{aligned}
\ee 
where the charges $(E,j_1,j_2,R,r)$ and $f_j$ generate the Cartan subalgebras of the superconformal and flavor symmetry groups respectively. Here we assume $|q|<1$ whereas $|a_j|=1$.

When the superconformal field theory in question admits a weakly coupled Lagrangian description, the superconformal index can be computed from the path integral definition. An important example is a free hypermultiplet in the bifundamental representation of $\mbox{SU}(N)\times\mbox{SU}(N)$
\be
\begin{aligned}
\label{bifindex}
\cI_B(a_i,b_j,c) & = \mbox{PE}\left[ \; \sum_{i,j=1}^N \frac{\sqrt{q}}{1-q}\left( a_i b_j c+\frac{1}{a_i b_j c}\right) \right] \\
& = \prod_{i,j=1}^{N}\left[ \; \prod_{m=0}^{\infty} \frac{1}{1-q^{\frac{1}{2}+m}a_i b_j c} \frac{1}{1-q^{\frac{1}{2}+m}(a_i b_j c)^{-1}}\right] \, ,
\end{aligned}
\ee
where $
\prod_{i=1}^N a_i = \prod_{j=1}^N b_j = 1
$
and $c$ is the fugacity for an additional U$(1)$ flavor symmetry. When $N=2$ the additional U$(1)$ flavor symmetry is enhanced to SU$(2)$ and we have an equivalent description as a half-hypermultiplet in the trifundamental of an SU$(2)^3$ flavor symmetry.

Furthermore, given the index $\cI(a_i,\ldots)$ of a superconformal field theory with an $SU(N)$ flavor symmetry, the index of the theory where this symmetry has been gauged is
\be
\int [da] \, \cI_V(a_i)\, \cI(a_i,\ldots)
\ee
where
\be
\begin{aligned}
\cI_V(a_i) & = \mbox{PE}\left[ -\frac{2q}{1-q}\left(\sum_{i,j=1}^N a_i/a_j-1\right)\right] \\ 
\end{aligned}
\ee
is the index of an SU$(N)$ vectormultiplet and
\be
[da]= \frac{1}{N!}\prod_{i=1}^{N-1}\frac{da_i}{2\pi i a_i} \prod_{i\neq j}^N\left(1-\frac{a_i}{a_j}\right)
\ee
is the Haar measure on SU$(N)$. In this manner, the index of large classes of superconformal field theories can be built from elementary building blocks.

Here we consider $\cN=2$ superconformal theories of class $\cS$ with maximal SU$(N)$ flavor symmetries~\cite{Gaiotto:2009we,Gaiotto:2009hg}, obtained by compactifying $N$ M5 branes on a Riemann surface $C$ with maximal punctures. The building block is a non-conventional superconformal field theory $\cT_N$ with SU$(N)^3$ flavor symmetry, associated to the sphere with three punctures. For each pants decomposition of the Riemann surface $C$, there is a description of the corresponding superconformal field theory obtained by taking a copy of $\cT_N$ for each pair-of-pants and gauging diagonal combinations of SU$(N)$ flavor symmetries. 

In general, the index of $\cT_N$ cannot be computed directly, since the relevant theory does not have a Lagrangian description. For the particular case of $N=3$, this index was computed in \cite{Gadde:2010te} by embedding it into a larger theory which admits a Lagrangian description in a different S-duality frame.  In \cite{Gadde:2011ik},  the relation to 2d q-YM was shown for $N=2,3$ and then used to conjecture an expression for the index of more general  trinion theories $\cT_N$. These conjectures were then proven by exploiting the properties of the superconformal index in the presence of surface defects~\cite{Gaiotto:2012xa}.

\subsection{Vortices and surface defects}

\begin{figure}[htp]
\centering
\includegraphics{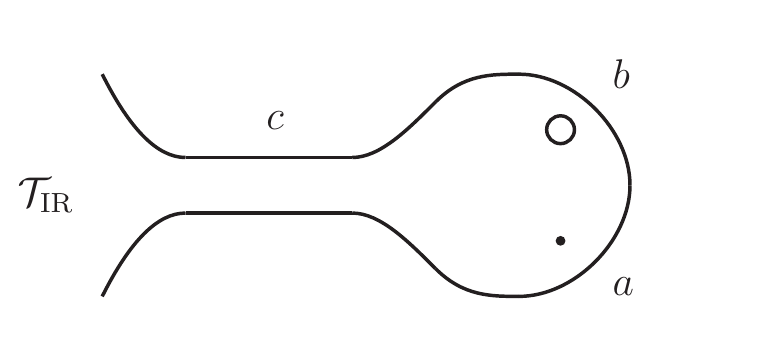}
\caption{The SCFT $\cT_{\mathrm{IR}}$, which is assumed to have only full punctures in this paper, is coupled to the SCFT associated to a sphere with three punctures (two maximal punctures and one minimal puncture, black dot in the picture). The resulting SCFT is called $\cT_{\mathrm{UV}}$ and corresponds to the degeneration limit shown in the figure.}
\label{figure:coupling}
\end{figure}

We now briefly review the construction of~\cite{Gaiotto:2012xa} for computing the superconformal index in the presence of surface defects. The starting point is a superconformal field theory $\cT_{\mathrm{IR}}$ with global flavor symmetry $\SU(N)$. This theory is then coupled to a hypermultiplet in the bifundamental representation of $\SU(N)\times\SU(N)$ by gauging a diagonal combination of the flavor symmetries - see figure~\ref{figure:coupling}. The resulting theory $\cT_{\mathrm{UV}}$ has an additional U$(1)_f$ flavor symmetry acting on the bifundamental hypermultiplet.

The two theories are related by a renormalization group flow $\cT_{\mathrm{UV}}\to \cT_{\mathrm{IR}}$, initiated by turning on a Higgs branch vacuum expectation value for the bifundamental hypermultiplet scalar $Q$. This can be implemented concretely at the level of the index. Let us denote the index of the original superconformal field theory $\cT_{\mathrm{IR}}$ by
\be
\cI(c_i,d_j,\ldots)\, .
\ee 
Then the index of $\cT_{\uv}$ is
\be
\cI_{\mathrm{UV}}(a,b_i,\ldots) = \int\, [\rd c] \; \cI_B(a,b_i,c_j) \; \cI_V(c_j) \;  \cI(c_j^{-1},d_k,\ldots)\, ,
\ee
where $a$ is the fugacity for the additional U$(1)_f$. The index of $\cT_{\uv}$ has simple poles coming from the chiral ring generated by $B=\det{Q}$ and its holomorphic derivatives in the plane rotated by $j=j_2-j_1$. Computing the residue at $a=q^{1/2}$ corresponds to a constant vacuum expectation value and leads back to $\cT_{\ir}$. However, computing the residue at points $a=q^{1/2+r/N}$ corresponds to a background vortex configuration of winding number $r\in\mathbb{Z}_{\geq0}$. It can be shown that computing the residue is equivalent to a difference operator acting on the original index

\be
\begin{aligned}
G_r \cdot \cI(b_i) &=  N \, \mathrm{PE} \left[-\frac{2q}{1-q} \right] \,  \underset{a=q^{1/2+r/N}}{\mathrm{Res}}   \, \frac{1}{a} \cI_{\uv}(a,b_i,\ldots)  = \\[+5pt]
 & = \sum_{\sum_{i=1}^N n_i = r} \left[ \, \prod_{i,j=1}^N \prod_{n=0}^{n_i-1} \frac{1-q^{1+n_j-n_i+n}b_i/b_j}{1-q^{n-n_j}b_j/b_i} \right]\cI (q^{r/N-n_i}b_i)\, ,
 \end{aligned}
 \label{Gdef}
\ee
which is a specialization to the Schur limit of the result found in~\cite{Gaiotto:2012xa}. This difference operator is interpreted as introducing a supersymmetric surface defect into the theory $\cT_{\ir}$ coupled to the SU$(N)$ flavor symmetry and labelled by $r\in\mathbb{Z}_{\geq0}$. The case $r=0$ corresponds to no surface operator.

The difference operators $G_{r}$ are commuting and self-adjoint in the propagator measure. They have common orthonormal eigenfunctions $\{\psi_{\cS}(a_i)\}$ labelled by irreducible representations ${\cal S}$ of SU$(N)$. The eigenfunctions are
\be
\psi_{\cS}(a_i) = \chi_{\cS}(a_i) / \sqrt{\cI_V(a_i)}\, ,
\ee
where 
\be
\chi_{\cS}(a_i) = \frac{\det (a_i^{s_j+N-j})}{\det(a_i^{N-j})} 
\ee
are the Schur polynomials of $\SU(N)$. Here we are labelling irreducible representations by partitions $\cS=(s_1,\ldots,s_{N-1},0)$, or equivalently, by the lengths of the rows of the corresponding Young diagram. 

In order to compute the eigenvalues, we expand the bifundamental hypermultiplet index in terms of the eigenfunctions
\be
I_B(a,b_i,c_j) = \sum_{\cS} \, \phi_{\cS}(a) \, \psi_{\cS}(b_i) \, \psi_{\cS}(c_j) 
\ee
where
\be
\phi_S(a) \, \propto \, \mathrm{PE} \left[ \frac{q^{N/2}}{1-q}(a^N+a^{-N} )\right]\chi_{\cS}(a q^{\frac{N-2}{2}},\ldots,aq^{-\frac{N-2}{2}},a^{1-N}) 
\ee
is the wavefunction for the U$(1)$ puncture, given here up to $a$-independent factors. Now, using the residue definition~\eqref{Gdef} of the operator $G_r$ and acting on the wavefunction $\psi_{\cS}(b_i)$, we find
\be
\begin{aligned}
E_r^{(\cS)} & = \underset{a=q^{r/N+1/2}}{\mathrm{Res}}\, \frac{1}{a} \phi_{\cS}(a) \; / \; \underset{a=q^{1/2}}{\mathrm{Res}}\, \frac{1}{a} \phi_{\cS}(a) \\[+5pt]
& = \left[ \prod_{j=0}^{r-1} \frac{1-q^{j+N}}{1-q^{-1-j}} \right] \frac{\chi_S(q^{\rho_1+\frac{r}{N}},\ldots,q^{\rho_{N-1}+\frac{r}{N} },q^{\rho_N+\frac{r}{N}-s})}{\chi_S(q^{\rho_1},\ldots,q^{\rho_N})} \, ,
\end{aligned}
\ee
where $\rho=(\rho_1,\ldots,\rho_N)$ is the Weyl vector in the orthogonal basis. One can show that this expression reduces to

\be
E_r^{(\cS)} = (-1)^r q^{r(r+N)/2}\frac{S_{\cS,\cR}}{S_{\cS,0}}\, ,
\ee
where $\cR=(r,0,\ldots,0)$ is the symmetric tensor representation of $\SU(N)$ and $S_{\cR,\cS}$ is an analytic continuation of the modular S matrix of the $\SU(N)$ WZW model away from integer level - see Appendix A. It is straightforward to derive this expression for the $A_1$ case, as done in \cite{Gaiotto:2012xa}, but the general $A_{N-1}$ case is somewhat more involved. This formula strongly suggests that the operators $G_r$ constructed above are associated to the symmetric tensor representations of $\SU(N)$, as claimed by \cite{Gaiotto:2012xa}.


\section{General difference operators}

The operators $G_r$ reviewed above can be associated to symmetric tensor representations $\cR=(r,0,\ldots,0)$ of $\SU(N)$. In this section, we construct surface defects associated to general irreducible representations $\cR=(r_1,\ldots,r_{N-1},0)$. The idea is to start from known difference operators and generate new ones by considering compositions $G_r \cdot G_{r'}$. Before proceeding let us consider the following transformation
\be
\label{simil}
 \tilde{G}_{r} \, \equiv \,  (-1)^{r} q^{-\frac{1}{2}r(r+N)}\, \cI_V^{-1/2}(a) \cdot G_{r} \cdot \cI_V^{1/2}(a)\, ,
\ee
which has the following action
\be
\tilde{G}_{r} \cdot \cI(a_i) = q^{-\frac{1}{2}r(N-1)} \sum_{\sum_{i=1}^N n_i = r} \, \left[ \, \prod_{i<j} \frac{q^{n_j}a_i-q^{n_i}a_j}{a_i-a_j} \right] \,  \cI(q^{r/N-n_i}a_i)\, .
\ee
The transformed operators $\tilde{G}_r$ have the following properties:
\begin{enumerate}
\item They are a commuting set of operators.
\item They are self-adjoint with respect to the Haar measure.
\item Their eigenfunctions are Schur polynomials $\chi_{\cS}(a_j)$.
\item Their eigenvalues are $\,S_{\cS,\cR}/S_{\cS,0}\,$ with $\cR=(r,0,\ldots,0)$.
\end{enumerate}
The transformation \eqref{simil} was designed so that the transformed operators are self-adjoint with respect to the Haar measure and their eigenvalues are ratios of S matrices, with no additional factors.

Let us start by reconsidering the superconformal index with surface defects for $A_1$ theories. The surface defects are labelled by $r\in\mathbb{Z}_{\geq0}$ corresponding to the irreducible representation of $\SU(2)$ of dimension $(r+1)$. The transformed difference operator becomes
\be
\tilde{G}_r \cdot \cI(a) = q^{-\frac{r}{2}}\, \sum_{n_1+n_2=r}  \frac{q^{n_2}a-q^{n_1}a^{-1}}{a-a^{-1}}\, \cI(q^{\frac{r}{2}-n_1}a)\, .
\ee
For $A_1$ theories, this exhausts the irreducible representations and so we expect the algebra to close. Indeed, it is straightforward to check that the composition of the operators decomposes according to the tensor product of their representations
\be
\begin{aligned}
\tilde{G}_{r_1} \cdot \tilde{G}_{r_2}  = \sum_{r=0}^{\infty} \, {\cN_{r_1,r_2}}^r \, \tilde{G}_r  = \sum_{r=|r_1-r_2|}^{r_1+r_2} \tilde{G}_r\, .
\end{aligned}
\ee
The numbers ${\cN_{r_1,r_2}}^{r}$ are the Littlewood-Richardson coefficients. This is equivalent to the formula
\be
S_{s,r_1} \, S_{s,r_2}  = S_{s,0} \sum_{r=0}^{\infty}\, {\cN_{r_1,r_2}}^{r} \, S_{s,r}  
\ee
relating their eigenvalues, where we have analytically continued the modular S matrix in $q$ away from rational points on the unit circle. Consequently, there is no truncation of representations.

As expected, for the $A_1$ case, we did not generate new difference operators. In the following we will study the composition of operators for the higher rank gauge groups, focusing in cases of increasing complexity.

\subsection{Closing the algebra for $SU(3)$}

The transformed difference operators corresponding to symmetric tensor representation $\cR=(r,0,0)$ of $SU(3)$  become
\be
\tilde{G}_{r} \cdot \cI(a_i) = q^{-r} \sum_{\sum_{i=1}^3n_i = r} \, \left[ \, \prod_{i<j} \frac{q^{n_j}a_i-q^{n_i}a_j}{a_i-a_j} \right] \,  \cI(q^{r/3-n_i}a_i)\, .
\label{conjDOsu3}
\ee
Under composition, we find that the algebra is not closed and we generate difference operators associated to more general representations $\cR=(r_1,r_2,0)$. In order to construct such operators, we will require
\be
\tilde{G}_{\cR_1} \cdot \tilde{G}_{\cR_2} = {\cN_{\cR_1,\cR_2}}^{\cR} \, \tilde{G}_{\cR}\, ,
\ee
where ${\cN_{\cR_1,\cR_2}}^{\cR}$ are the Littlewood-Richardson coefficients. In other words, the operators should decompose as the tensor product of their  representations. We proceed constructively.

First, consider the difference operator for the fundamental representation $\mathbf{3}$. This is given by
\be
\begin{aligned}
\tilde{G}_{\tiny\yng(1)} \cdot \cI(a_1,a_2,a_3) & = 
\frac{\left(a_1-a_2 q\right) \left(a_1-a_3 q\right)}{\left(a_1-a_2\right) \left(a_1-a_3\right) q} \, \cI \left(q^{-\frac{2}{3}}a_1,q^{\frac{1}{3}}a_2,q^{\frac{1}{3}}a_3\right) \\[10pt]
& + \frac{\left(a_1 q-a_2\right) \left(a_2-a_3 q\right) }{\left(a_1-a_2\right) \left(a_2-a_3\right) q} \, \cI \left(q^{\frac{1}{3}}a_1,q^{-\frac{2}{3}}a_2,q^{\frac{1}{3}}a_3\right)\\[10pt]
& + \frac{\left(a_1 q-a_3\right) \left(a_2 q-a_3\right) }{\left(a_1-a_3\right) \left(a_2-a_3\right) q}\, \cI \left(q^{\frac{1}{3}}a_1,q^{\frac{1}{3}}a_2,q^{-\frac{2}{3}}a_3 \right) \, . \\[10pt]
\end{aligned}
\ee
The three terms are related by cyclic symmetry in $(a_1,a_2,a_3)$. We interpret each of them as associated to a state in the fundamental representation, that is, to the \emph{semi-standard Young tableaux} (see Appendix~\ref{appendix:smatrix} for our notation)
\be
\young(1) \quad (1,0,0)  \qquad \young(2) \quad (0,1,0)  \qquad \young(3) \quad (0,0,1)\, .
\ee
Here we have introduced the notation $(n_1,n_2,n_3)$, where $n_i$ denotes the number of times that $i$ appears in the tableau. The sum over tableaux is therefore equivalent to the sum over all $n_i$ such that $\sum_{i=1}^3n_i=1$.

 Similarly, for the symmetric tensor $\mathbf{6}$ we have the following Young tableaux
\be
\begin{aligned}
& \young(11) \quad (2,0,0) \qquad \young(12) \quad (1,1,0)\\
& \young(13) \quad (1,0,1) \qquad \young(22) \quad (0,2,0)\\
& \young(23) \quad (0,1,1) \qquad \young(33) \quad (0,0,2)\, ,
\end{aligned}
\ee 
reproducing the sum $\sum_{i=1}^3n_i=2$. More generally, for any symmetric tensor representation $\cR=(r,0,0)$, the sum over all $n_i$ such that $\sum_{i=1}^3n_i =r$ is equivalent to summing over semi-standard Young tableau.

Now compose two operators in the fundamental representation. From the tensor product rule $\mathbf{3}\times\mathbf{3}=\mathbf{6}+\mathbf{3}^*$ and our knowledge of the operator for the symmetric tensor $\mathbf{6}$, we can construct an operator for the antisymmetric tensor $\mathbf{3}^*$. In this way, we find
\be
\begin{aligned}
\tilde{G}_{\tiny\yng(1,1)} \cdot \cI(a_1,a_2,a_3)  & \, = \,
\frac{\left(a_1 q-a_2\right) \left(a_1 q-a_3\right) }{\left(a_1-a_2\right) \left(a_1-a_3\right)  q}\, \cI \left(q^{\frac{2}{3}}a_1,q^{-\frac{1}{3}}a_2,q^{-\frac{1}{3}}a_3\right) \\[+10pt]
& \, + \, \frac{ \left(a_1-a_2 q\right) \left(a_2 q-a_3\right) }{\left(a_1-a_2\right)  \left(a_2-a_3\right) q} \, \cI \left(q^{-\frac{1}{3}}a_1,q^{\frac{2}{3}}a_2,q^{-\frac{1}{3}}a_3\right) \\[+10pt]
& \, + \, \frac{ \left(a_1-a_3 q\right) \left(a_2-a_3 q\right) }{ \left(a_1-a_3\right) \left(a_2-a_3\right) q} \, \cI \left(q^{-\frac{1}{3}}a_1,q^{-\frac{1}{3}}a_2,q^{\frac{2}{3}}a_3\right) \, .\\[+10pt]
\end{aligned}
\ee
This operator is also obtained by summing the expression
\be
 q^{-2} \left[ \, \prod_{i<j} \frac{q^{n_j}a_i-q^{n_i}a_j}{a_i-a_j} \right] \,  \cI(q^{\frac{2}{3}-n_i}a_i)
\ee
over the tableaux
\be
\young(1,2) \quad (1,1,0) \qquad \young(2,3) \quad (1,0,1) \qquad \young(1,3) \quad (0,1,1)
\ee
for the antisymmetric tensor representation $\mathbf{3}^*$. Furthermore, we can explicitly check that this operator commutes with the difference operators already known, which is a consistency check of our construction. Note that this operator is related to that of the fundamental by $q\to q^{-1}$. 

Next, using the decomposition rule $\mathbf{3}\times\mathbf{6}=\mathbf{8}+\mathbf{10}$ and our knowledge of the operator for the symmetric tensor $\mathbf{10}$, we can derive an expression for the difference operator in the adjoint representation $\mathbf{8}$. We find an operator with eight terms, obtained by summing 
\be
 q^{-3} \left[ \, \prod_{i<j} \frac{q^{n_j}a_i-q^{n_i}a_j}{a_i-a_j} \right] \,  \cI(q^{1-n_i}a_i)
\ee
over the tableaux
\be
\begin{aligned}
& \young(11,2) \quad (2,1,0) \qquad \young(12,2) \quad (1,2,0)\\
& \young(11,3) \quad (2,0,1) \qquad \young(12,3) \quad (1,1,1)\\
& \young(13,2) \quad (1,1,1) \qquad \young(13,3) \quad (1,0,2)\\
& \young(22,3) \quad (0,2,1) \qquad \young(23,3) \quad (0,1,2)\; .
\end{aligned}
\ee 
There is now an important consistency check. Using our derived expression for the operator in the antisymmetric tensor $\mathbf{3}^*$ and the tensor product rule $\mathbf{3}\times\mathbf{3}^*=\mathbf{1}+\mathbf{8}$, we can find a second expression for the operator in the adjoint representation. We find that the two constructions agree precisely.

The above examples are sufficient to state our proposal for the difference operator associated to a general irreducible representation $\cR=(r_1,r_2,0)$. Denoting the set of semi-standard Young tableaux associated to this representation by $\cY_{\cR}$, we propose that
\be
 \tilde{G}_{\cR} \cdot \cI(a_i) = q^{-|\cR|} \sum_{\cY_{\cR}} \left[ \, \prod_{i<j} \frac{q^{n_j}a_i-q^{n_i}a_j}{a_i-a_j} \right] \,  \cI(q^{\frac{|\cR|}{3}-n_i}a_i)\, ,
\ee
where $|\cR|$ is the number of boxes in the tableau. Note that this expression depends only on the numbers $(n_1,n_2,n_3)$ associated to each Young tableau - some terms can contribute many times. We have extensively checked that this formula reproduces the operators obtained by our constructive approach. 

\subsection{Closing the algebra for $SU(N)$}

We can now state our conjecture for the difference operator labelled by any irreducible representation $\cR=(r_1,\ldots,r_{N-1},0)$ of SU$(N)$. As above, we denote the set of semi-standard tableaux by $\cY_{\cR}$ and assign them the labels $n_i$ encoding the number of times that $i=1,\ldots,N$ appears. Finally, $|\cR|$ is the number of boxes in the Young diagram. For any representation,
\be
\tilde{G}_{\cR} \cdot \cI(a_i) = q^{-\frac{1}{2}|\cR|(N-1)} \sum_{\cY_{\cR}} \, \left[ \, \prod_{i<j} \frac{q^{n_j}a_i-q^{n_i}a_j}{a_i-a_j} \right] \,  \cI(q^{\frac{|\cR|}{N}-n_i}a_i) \, .
\label{Gconj}
\ee
This family of operators $\{\tilde{G}_{\cR}\}$ has the following properties (see appendix B)
\begin{enumerate}
\item Self-adjoint with respect to the Haar measure.
\item Commutativity.
\item Eigenfunctions are Schur functions: $\chi_{\cS}(a_j)$.
\item Eigenvalues: $\,S_{\cS,\cR}/S_{\cS,0}\,$.
\item Closed under composition: $
\tilde{G}_{\cR_1} \cdot \tilde{G}_{\cR_2} = {\cN_{\cR_1\cR_2}}^{\cR} \, \tilde{G}_{\cR}
$
\end{enumerate}
The final two properties are actually equivalent due to the formula
\be
S_{\cS,\cR_1} \, S_{\cS,\cR_2}  = S_{\cS,0} \sum_{\cR=0}^{\infty}\, {\cN_{\cR_1,\cR_2}}^{\cR} \, S_{\cS,\cR}  
\ee
where ${\cN_{\cR_1,\cR_2}}^{\cR}$ are the Littlewood-Richardson coefficients. Finally, let us mention that given the representation ${\cal R}$ and its complex conjugate $\bar{\cal R}$, their difference operators are related by $q \leftrightarrow q^{-1}$.

\subsection{Back to the index}

Now reversing the transformation, we obtain the following operators acting on the superconformal index
\be
\begin{aligned}
G_{\cR} \cdot \cI(a_i) \; & = \sum_{\cY_{\cR}} \left[\, \prod_{i,j=1}^N  \prod _{n=0}^{n_i-1} \frac{1-a_i a_j^{-1} q^{n-n_i+n_j+1}}{1-a_j a_i^{-1}q^{n-n_j}} \right]\, \cI(q^{|\cR|/N-n_i}a_i)\\
& =(-1)^{|{\cal R}|} q^{\gamma({\cal R})} \sum_{\cY_{\cR}} \left[  \prod_{i=1}^N a_i^{Nn_i}\right] \, q^{-\frac{N-1}{2}\sum_i n_i^2 +\sum_{i<j}n_in_j}\, \cI(q^{|\cR|/N-n_i}a_i)\, .
\end{aligned}
\ee
They obey the recursion relation
\be
G_{\cR_1}\cdot G_{\cR_2} = \sum_{\cR}(-1)^{|\cR_1|+|\cR_2|-|\cR|} q^{\gamma(\cR_1)+\gamma(\cR_2)-\gamma(\cR)} {\cN_{\cR_1\cR_2}}^{\cR}G_{\cR}\; ,
\ee
where we have defined $\gamma(\cR) = |\cR|(|\cR|+N)/2$. Note that typically $|\cR_1|+|\cR_2|\neq|\cR|$ because columns of length $N$ are deleted from the Young diagrams created by the tensor product. It is natural to conjecture the existence of surface defects labelled by general representations $\cR$ and that the above operators compute the superconformal index in the presence of these defects.


\section{Surface defects and $\mathbf{q}$-deformed YM}

The superconformal index is invariant under continuous deformations of the superconformal field theory~\cite{Kinney:2005ej}. Invariance under generalized $S-$duality implies that the superconformal index of SU$(N)$ generalized quivers is computed by a topological quantum field theory on $C$  \cite{Gadde:2009kb}. In the Schur limit, the relevant topological quantum field theory is q-deformed YM with gauge group SU$(N)$ in the limit of vanishing area~\cite{Gadde:2011ik,Gadde:2011uv}. This 2d theory can be understood as an analytic continuation of Chern-Simons theory on $C \times S^1$.

In this section, we extend the dictionary between the superconformal index and $q$-deformed YM to include the surface defects labelled by any irreducible representation $\cR$. While the maximal SU$(N)$ flavor punctures correspond to fixing the holonomy of the connection around the punctures in $C$, we find that a surface defect corresponds to inserting a Wilson loop in the representation $\cR$ wrapping the $S^1$. The transformed difference operators constructed in the previous section, $\tilde{G}_{\lb}$, then provide a representation of the Verlinde algebra.

\subsection{q-deformed YM and holonomy punctures}

As already mentioned, q-deformed YM theory on $C$ in the zero area limit can be understood as an analytic continuation of Chern-Simons theory on $C\times S^1$ \cite{Aganagic:2004js}. The fundamental variables are the connection $A$ on $C$ and a periodic adjoint valued scalar $\phi$ given by the holonomy of the Chern-Simons connection around the $S^1$,
\be
e^{i \phi} = \mbox{P} \exp \left( i \oint_{S^1} A \right)\, .
\ee
The gauge fixed path integral is 
\be
Z \sim \int \prod d\phi_i \left( \Delta(\phi)\right)^{\chi(C)}\exp{\left(-\frac{1}{g_s} \int_{C} \sum_i \phi_i \, F_i \right)}\; ,
\ee
where the path integral measure $\Delta(\phi) = \prod_{1 \leq i < j \leq N} 2 \sin\left( \frac{\phi_i-\phi_j}{2}\right)$ takes into account the periodicity of $\phi$ and leads to the deformation with parameter $q=e^{-g_s}$. This provides an analytic continuation of Chern-Simons theory away from integer level $k$ by moving $q$ away from rational points $\e^{2\pi i/(k+N)}$ on the unit circle. 

The partition function on a Riemann surface $C$ with boundaries can be evaluated by surgery. The starting point for this construction is the Hilbert space obtained by Hamiltonian quantization on $\mathbb{R}\times S^1$. This is given by gauge invariant functions of the connection $A$, which are symmetric polynomials in the holonomy eigenvalues $ a=(a_1,\ldots,a_{N-1})$ of the connection around the $S^1$. The path integral on a Riemann surface with a boundary where the holonomy eigenvalues are held fixed at ${a}$ defines a wavefunction $\Psi({ a})$ in the Hilbert space associated to that boundary.

A convenient basis is given by the Schur polynomials $\chi_{\cS}(a)$ labelled by irreducible representations $\cS=(s_1,\dots,s_{N-1},0)$. The Schur polynomials are orthonormal in the Haar measure
\be
\int [d{a}] \, \chi_{\cS_1}({ a}) \, \chi_{\cS_2}({ a}^{-1}) = \delta_{\cS_1,\cS_2} \, ,
\ee
and any wavefunction can be expanded in terms of those
\be
\psi({ a}) = \sum_{\cS} \psi_{\cS} \, \chi_{\cS}({ a})\, ,
\qquad
\psi_{\cS} = \int [d{ a}] \, \chi_{\cS}({ a}) \, \psi({ a}^{-1})\, .
\ee
The amplitudes for Riemann surfaces with boundaries can be glued by identifying the holonomy eigenvalues and integrating with respect to the Haar measure.

\begin{figure}[htp]
\centering
\includegraphics{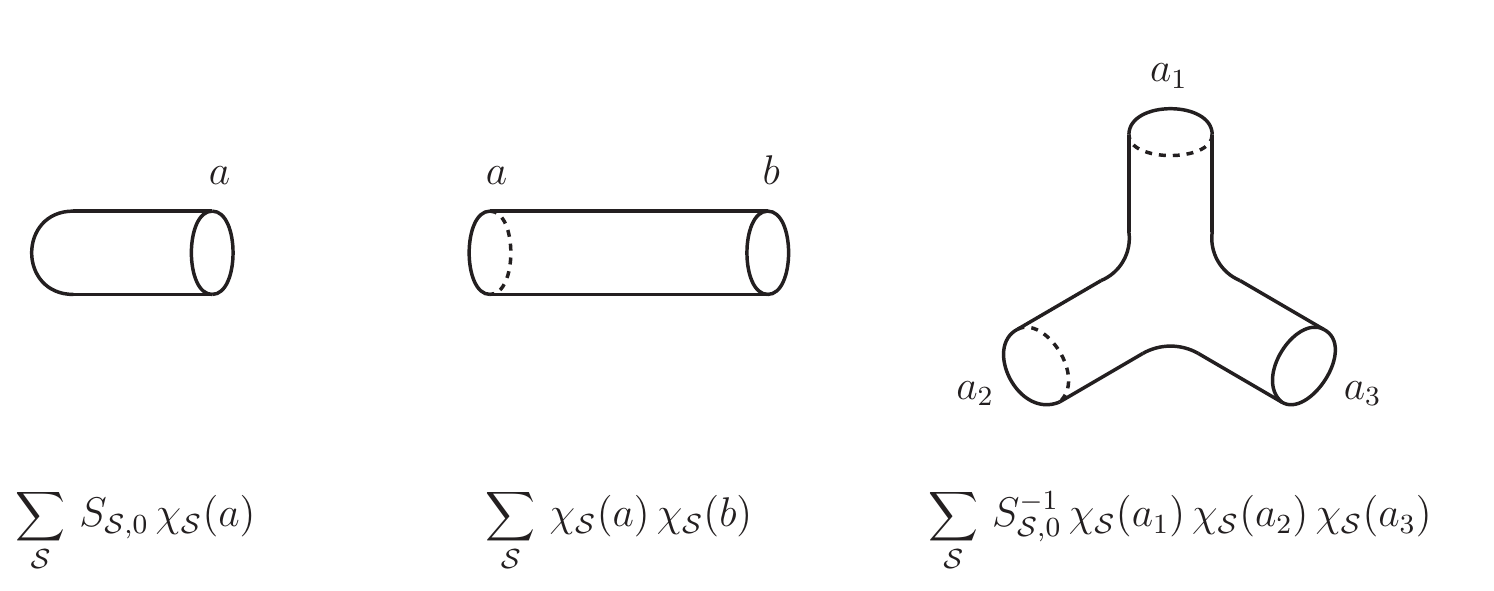}
\caption{Partition functions for the sphere with one, two and three holes/punctures respectively, for q-deformed YM in the zero area limit.}
\label{figure:basicamplitudes}
\end{figure}

The partition function on any Riemann surface with boundaries can be computed by gluing together the basic amplitudes with one, two and three boundaries, shown in figure~\ref{figure:basicamplitudes}. The general result for a Riemann surface of genus $g$ with $n$ punctures with boundary conditions fixed to ${a}_1,\dots,{a}_n$ is given by
\be
\label{fact}
\sum_{\cS} S_{\cS,0}^{2-2g-n} \,  \chi_{\cS}({ a}_1) \ldots \chi_{\cS}({ a}_n).
\ee
The rescaled Schur index $\tilde{\cI}_{g,n}({a}_1,...,{a}_n)$\footnote{For each flavor puncture, the physical index $\cI_{g,n}({a}_1,...,{a}_n)$ is multiplied by the square root of the corresponding vectormultiplet index, plus an overall factor which depends only on $q$.} associated to a genus $g$ theory with $n$ maximal punctures, was shown to agree exactly with this expression \cite{Gadde:2011ik,Gaiotto:2012xa}. 

\subsection{Adding defect punctures}

The partition function can also be enriched by the insertion of gauge invariant operators constructed from the scalar
\be
\cO_{\cR}\equiv \chi_{\cR}(e^{i\phi})\, ,
\ee 
whose correlators are independent of their position on $C$~\cite{Aganagic:2004js}. In Chern-Simons theory on $S^1\times C$, an insertion of the operator $\cO_{\cR}$ on $C$ corresponds to inserting a Wilson loop around the $S^1$ in the representation $\cR$. In the following, we will refer to insertions of such operators as `defect punctures'.

\begin{figure}[htp]
\centering
\includegraphics{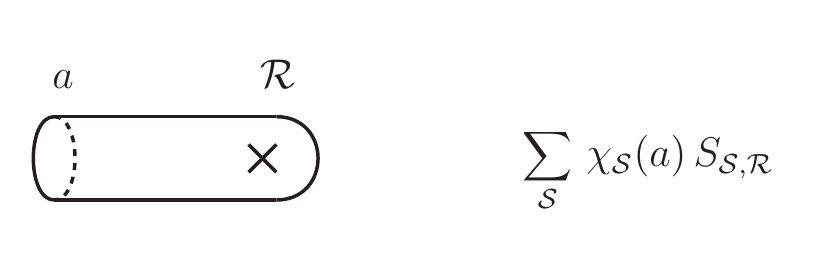}
\caption{Disk amplitude in the presence of a surface puncture.}
\label{figure:defectamplitude}
\end{figure}

To compute correlation functions with defect punctures, we return to the cylinder amplitude with holonomy eigenvalues $a$ and $b$ respectively - see figure~\ref{figure:basicamplitudes}. In the Chern-Simons theory on $S^1\times C$, the boundary with holonomy eigenvalues $b$ becomes a boundary torus $S^1\times \tilde{S}^1$. The first step is to interchange those circles using the modular S matrix
\be
\sum_{\cS}\,   \chi_{\cS}(a) \, \chi_{\cS}(b) \; \xrightarrow{~\tilde{S}^1 \leftrightarrow S^1~} \; 
 \sum_{\cS,\cS'}\,   \chi_{\cS}(a) \, S_{\cS,\cS'} \, \chi_{\cS'}(b) \, ,
\ee
so that we are now fixing the holonomy eigenvalues $b$ on $S^1$. To insert a Wilson line in the representation $\cR$, we multiply by $\chi_{\cR}(b)$ and integrate over the holonomy $b$. From the orthonormality of Schur polynomials with respect to the Haar measure we find
\be
\la \cO_\cR \ra_{0,1}  = \sum_{\cS}\,   \chi_{\cS}(a) \, S_{\cS,\cR} \, .
\ee
This is the amplitude for a disk with holonomy eigenvalues $a$ and a defect puncture labelled by the irreducible representation $\cR$ - see figure~\ref{figure:defectamplitude}.

Any amplitude with defect punctures can now be calculated by gluing the above amplitude to fixed holonomy boundaries. For example, starting from a sphere with four boundaries, we construct the amplitude with three boundaries and one defect puncture in the representation $\cR$
\be
\la \cO_\cR \ra_{0,3} =  \sum_{\cS}  \frac{ \chi_{\cS}(a_1) \chi_{\cS}(a_2) \chi_{\cS}(a_3) }{S_{\cS,0}}  \, \frac{S_{\cS,\cR}}{S_{\cS,0}} \, .
\ee
Similarly, starting from the sphere with three boundaries we can obtain the amplitude for three defect punctures in representations $\cR_1,\cR_2$ and $\cR_3$
\be
\begin{aligned}
 \la \cO_{\cR_1}\cO_{\cR_2}\cO_{\cR_3} \ra_{0,0} 
 & = \sum_{\cS} \, \frac{S_{\cS,\cR_1}\, S_{\cS,\cR_2}\, S_{\cS,\cR_3}}{S_{\cS,0}} \\
 & =  {\cN_{\cR_1,\cR_2}}^{\bar\cR_3}\, ,
\end{aligned}
\ee
where, since we are analytically continuing $q$ away from roots of unity, there is no truncation of representations and the numbers ${\cN_{\cR_1,\cR_2}}^{\cR_3}$ are simply the Littlewood-Richardson coefficients for SU$(N)$. This is an analytic continuation of the Verlinde formula.

In full generality, for the amplitude of a Riemann surface of genus $g$, with $n$ boundaries with fixed holonomies $(a_1,\ldots,a_n)$ and defect punctures in representations $(\cR_1,\ldots,\cR_l)$ we have
\be
\label{def}
\la \cO_{\cR_1} \ldots \cO_{\cR_l} \ra_{g,n}=  \sum_{\cS}  S_{\cS,0}^{2-2g-n}\, \prod_{i=1}^n \chi_{\cS}(a_i)  \, \prod_{j=1}^l \frac{S_{\cS,\cR_j}}{S_{\cS,0}}\, .
\ee
Therefore, adding a defect puncture labelled by the irreducible representation $\cR$ inserts a factor $S_{\cS,\cR}/S_{\cS,0}$ into the sum over representations. From the formula
\be
S_{\cS,\cR_1} \, S_{\cS,\cR_2}  = S_{\cS,0} \sum_{\cR=0}^{\infty}\, {\cN_{\cR_1,\cR_2}}^{\cR} \, S_{\cS,\cR}  
\ee
we derive the operator product expansion
\be
\label{OPE}
\cO_{\cR_1} \cdot \cO_{\cR_2} = \sum_{\cR} {\cN_{\cR_1\cR_2}}^{\cR} \, \cO_{\cR}
\ee
inside the correlation function. This is (an analytic continuation of) the representation of the Verlinde algebra for Chern-Simons theory on $S^1 \times C$.

Now we are ready to spell out the precise dictionary. First recall that the rescaled index without surface defects is given by the expression~\eqref{fact} involving a sum over irreducible representations $\cS$. When acting with the transformed difference operator $\tilde{G}_{\cR}$ on a flavor puncture, each term in the sum picks up a factor of $S_{{\cal S},\cR} / S_{\cS,0}$. This is equivalent to the insertion of a defect puncture ${\cal O}_{\cR}$ in the $q$-deformed YM correlator, see  (\ref{def}). Hence, surface defects in the 4d theory correspond to defect punctures in $q$-deformed YM, see figure 4. Finally, let us mention that the OPE expansion (\ref{OPE}) guarantees consistency if we add several defect punctures.

\begin{figure}[htp]
\centering
\includegraphics{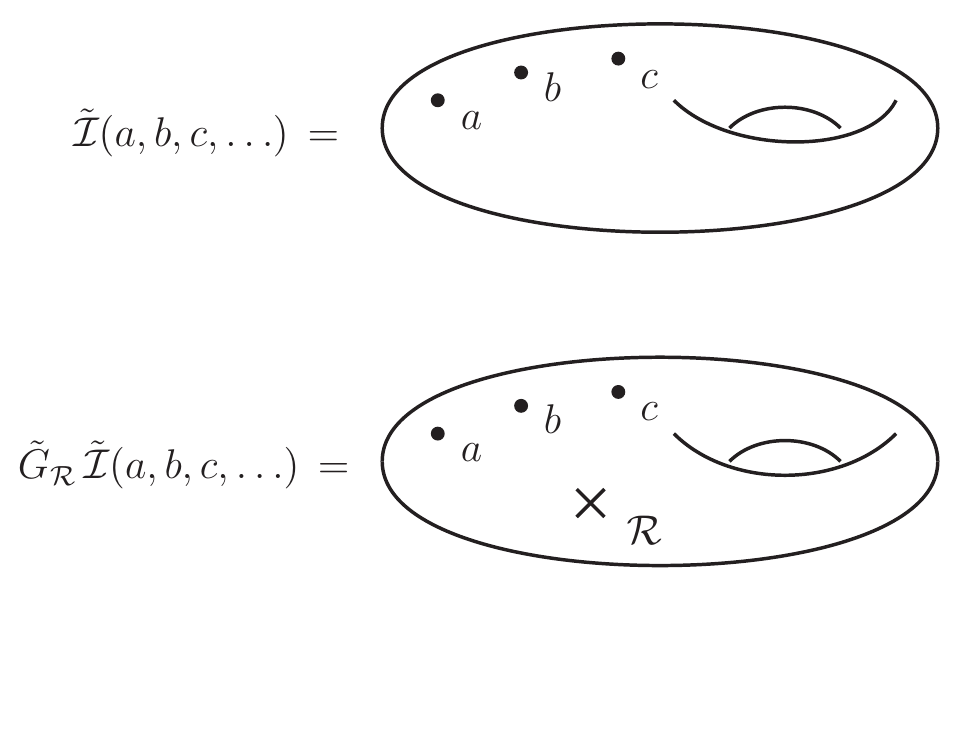}
\caption{The index without surface defects corresponds to a correlator of q-YM. Inserting a surface defect/acting with a difference operator in the 4d theory corresponds to inserting a defect puncture on the 2d side.}
\label{figure:dictionary}
\end{figure}

\section{Conclusions}

In this paper we have considered the superconformal index of 4d ${\cal N}=2$ theories of type $A_{N-1}$ in the presence of certain surface defects labelled by arbitrary irreducible representations of $SU(N)$. This can be obtained by applying certain difference operators to the index without defects. These operators can be constructed from the ones given in \cite{Gaiotto:2012xa}, which are labelled by symmetric representations, by closing the algebra under composition. We restrict to a one-parameter family in the fugacity space $(p,q,t) = (0,q,q)$ called the Schur limit. After a similarity transformation the operators have the following properties:
\begin{enumerate}
\item Self-adjoint in the Haar measure.
\item Commutativity: $\quad \left[ \,\tilde{G}_{\cR_1},\tilde{G}_{\cR_2} \right] = 0$ 
\item Composition: $\quad \tilde{G}_{\cR_1} \cdot \tilde{G}_{\cR_2} = {\cN_{\cR_1\cR_2}}^{\cR} \, \tilde{G}_{\cR}$
\item Eigenvalues: $\quad \tilde{G}_{\cR} \; \chi_{\cS}(a) = \frac{S_{\cR,\cS}}{S_{0,\cS}}\, \chi_{\cS}(a)$
\end{enumerate}
where ${\cN_{\cR_1\cR_2}}^{\cR}$ are the Littlewood-Richardson coefficients  and $S_{\cR,\cS}$ is an analytic continuation of the modular S matrix of the SU$(N)$ WZW model. Furthermore, $\tilde{G}_{\cR}$ is given by a sum of terms, were each term corresponds to the semi-standard Young tableau in the representation $\tilde{G}_{\cR}$.

Further we showed that, from the 2d perspective, acting with such difference operators is equivalent to the insertion of a gauge invariant operator  ${\cal O}_\cR = \chi_\cR(e^{i\phi})$ constructed from the scalar field of q-YM. This makes precise the dictionary suggested in \cite{Gaiotto:2012xa}.

The surface defects discussed in this paper should descend from codimension-four defects in the 6d theory, located at a point in the Riemann surface $C$. These surface defects have not been much studied and we hope that the results of this paper will contribute to understanding them more fully. There are several open problems. The most natural is the generalization of what we did to generic superconformal fugacities $(p,q,t)$. The algebra under composition should be much more complicated in this case, but should give information regarding the ``OPE coefficient theory" mentioned in  \cite{Gaiotto:2012xa}. Furthermore, it would be interesting to incorporate non-maximal punctures to our picture. Finally, it would be interesting to identify these surface defects from a purely 4d perspective, and to re-obtain the results (in the $A_1$ case) from a bona-fide localization computation. 

\acknowledgments

We would like to thank Davide Gaiotto and Paul Richmond for interesting discussions. The work of L.F.A. and M.F. is supported by ERC STG grant 306260. L.F.A. is a Wolfson Royal Society Research Merit Award holder. The work of M.B. is supported by EPSRC grant EP/J019518/1. The work of LH is supported by a NWO Rubicon fellowship and by NSF grant PHY-0757647. This work is in addition supported in part by the DOE grant DE-FG03-92-ER40701.

\appendix

\section{Group theory and modular S matrix}

\label{appendix:S matrix}

Finite dimensional representations of $SU(N)$ are in one to one correspondence with Dynkin labels $\lambda=(\lambda_1,\lambda_2,...,\lambda_{N-1})$, where $\lambda_i$ are non-negative integers. Equivalently we can label a representation ${\cal R}$ in terms of a partition $\ell_1 \geq \ell_2 \geq... \geq \ell_N=0$, where 

\begin{equation}
\ell_i = \lambda_i + \lambda_{i+1}+...+\lambda_{N-1} \, .
\end{equation}
To a partition we associate a Young tableau. For instance,  the following tableau

$$\yng(4,2,2)$$
corresponds to the partition $\{ 4,2,2\}$. If the partition $\{ \ell_1,...,\ell_k\}$ corresponds to a given representation ${\cal R}$, then the partition $\{ N- \ell_k,...,N-\ell_1\}$ corresponds to the {\it complex conjugate} representation $\bar {\cal R}$ 

Below we will find it useful to use the so called orthonormal $\epsilon$ basis. In this basis one obtains
\begin{equation}
\lambda = \sum_i^N (\ell_i-\kappa)\epsilon_i,~~~~~\kappa = \frac{1}{N} \sum_{j=1}^{N-1}j \lambda_j = \frac{1}{N} \sum_{j=1}^{N-1}j (\ell_j - \ell_{j+1})\, .
\end{equation} 

We can describe all the states in a given representation by using {\it semi-standard tableaux}. This involves filling the boxes of a Young tableau with positive integers. If $c_{i,j}$ is the integer appearing on the i-th row from top and the j-th column from left, then the following conditions should be satisfied
\begin{equation}
1 \leq c_{i,j} \leq N,~~~~~c_{i,j} \leq c_{i,j+1},~~~~~c_{i,j}<c_{i+1,j}\, ,
\end{equation}
namely, the numbers are non-decreasing from left to right and strictly increasing from top to bottom. Finally, to each semi-standard tableau we attach the labels $(n_1,...,n_N)$, where $n_i$ denotes the number of times that $i$ appears in the semi-standard tableau. As an example below we include a few semi-standard tableaux  for the ${\bf 8}$ representation of $SU(3)$ with their corresponding labels.

\begin{eqnarray*}
&\young(11,2) ~~~~~\young(12,3)~~~~~&\young(13,2) ~~~~~~ \young(22,3)  \\
&(2,1,0)~~~~~(1,1,1)~~~~~&(1,1,1)~~~~~(0,2,1)
\end{eqnarray*}

\subsection{Schur functions and modular S-matrix}

Given $a_i$, $i=1,...,N$ with $\prod_{i=1}^N a_i=1$, the Schur function (namely, the character in the orthonormal basis) of the representation ${\cal R}$ labelled  by the partition $\ell_i$ is given by
\begin{equation}
\chi_{\cal R}({\bf a}) = \frac{\det a_j^{\ell_i+N-i}}{\det a_j^{N-i}}\, .
\end{equation}
The {\it Weyl vector} is half the sum of the positive roots
\begin{equation}
\rho = \frac{1}{2} \sum_{\alpha \epsilon \Delta_+} \alpha\, .
\end{equation}

One can explicitly check that all Dynkin labels for this vector are equal to one. In the orthogonal basis mentioned above we have
\begin{equation}
\rho = \frac{1}{2} ( N-1,N-3,...,1-N)\, .
\end{equation}
We denote each element of this vector as $\rho_i$. Furthermore, given a partition $\{\ell_1,...,\ell_N \}$ we call $\kappa_i$ the corresponding elements of this vector in the orthogonal basis, namely

\begin{equation}
\kappa_i = \ell_i - \frac{1}{N} \sum_{j=1}^{N-1} j (\ell_j -\ell_{j+1}),~~~~i=1,...,N.
\end{equation}

Given two representations ${\cal R}$ and ${\cal R}'$ labelled by partitions $\ell_i$ and $\ell_i'$ the modular S-matrix is given by
\begin{equation}
S_{\bar {\cal R} , {\cal R}'} = S_{00} \chi_{\cal R}( q^{\rho_1},...,q^{\rho_N})\chi_{{\cal R}'}( q^{\rho_1+\kappa_1},...,q^{\rho_N+\kappa_N}) \, .
\end{equation}

\section{Relation to Macdonald operators}
\label{appendix:macdonal}

The transformed difference operators, as defined in the body of the text \eqref{Gconj},
are closely related to what are called q-difference operators in the mathematics literature \cite{MR1354144}. The latter 
constitute a commuting family of operators defined for general root systems. For a root system of type $A_{N-1}$,
the q-difference operators act on the space of symmetric polynomials with two
parameters, $(q,t)$, and they are given by \cite{MR1354144}:

\begin{equation}
 \mathcal{D}_{r}^{(N)} = t^{r (r-1)/2} \sum_{\substack{I\subset \{1,\ldots, N\}
\\ |I| = r} } \prod_{\substack{i\in I \\ j \notin I}} \frac{t x_i - x_j}{x_i -
x_j} \prod_{i \in I} \mathcal{T}_{q,x_i} \ , \quad  0 \leq r \leq N 
\label{qdifferenceoperator}
\end{equation}
where $\mathcal{T}_{q,x_i}$ is defined as
$\mathcal{T}_{q,x_i} \left[f \left( x_1, \ldots, x_N \right) \right] =
f \left( x_1, \ldots , x_{i-1}, q x_i, x_{i+1}, \ldots , x_N \right)$
for any polynomial $f$. By $|I|=r$ we mean that the subset $I \subset \{ 1,
\ldots, n \}$ is of cardinality $r$. Notice that $\mathcal{D}_{0}^{(N)} = 1$ and
$\mathcal{D}_{N}^{N} f(\{x_i\}) \propto f(\{ q x_i \})$.

These operators have been studied by mathematicians in order to prove certain properties
of the Macdonald functions for general root systems. These proofs use the fact that the
Macdonald functions (and hence the Schur functions in the $q=t$ limit) are common
eigenfunctions of the family $\{ \mathcal{D}_{r}^{(N)} \}$, for $A_{N-1}$ and similarly for more general root systems. 

It turns out that there is a very simple relationship
between these $q-$difference operators and our difference operators. Given \eqref{qdifferenceoperator}, we can express our difference operators
associated to antisymmetric tensors, $\mathcal{R}^{a}_r$, with purely vertical Young diagrams consisting of $r$ boxes,
$\tilde{G}_{\mathcal{R}^{a}_{r}}$, in terms of the q-difference operators:

\begin{equation}
\left[ \tilde{G}_{\mathcal{R}^{a}_{r}} \right] f\left( \left\{ a_i
\right\}_{i=1}^{N} \right) =q^{ r (N-1)/2} \left[
\hat{\mathcal{D}}_{r}^{(N)}\right]  f\left( \left\{ q^{r/N} a_i
\right\}_{i=1}^{N} \right) \label{surfopvsqdiff}\\
\end{equation} 
where we have adopted the Schur limit $q=t$. Additionally, we redefined the $\mathcal{D}^{(N)}_{r}$ operators by taking $q$ to
$q^{-1}$:

\begin{equation}
 \hat{\mathcal{D}}_{\ell}^{(N)} := \mathcal{D}_{\ell}^{(N)} \mid_{(t,q) \rightarrow (q^{-1},q^{-1})} = q^{-\ell (\ell-1)/2} \sum_{\substack{I\subset \{1,\ldots,
n\} \\ |I| = \ell} } \prod_{\substack{i\in I \\ j \notin I}} \frac{q^{-1} x_i -
x_j}{x_i - x_j} \prod_{i \in I} \mathcal{T}_{q^{-1},x_i}
\end{equation}
This does however not affect any of the properties of $\mathcal{D}_{r}^{(N)}$.
Therefore, as proven in \cite{MR1354144}, the operators 
$\hat{D}_{\ell}^{(N)}$ are commuting, self-adjoint with respect to the SU(N)
Haar measure and have the Macdonald polynomials as eigenfunctions. Hence, as a
result of the above identification \eqref{surfopvsqdiff}, and the fact that we can
construct $\tilde{G}_{\mathcal{R}}$ for any representation $\mathcal{R}$ from the fully antisymmetric ones, $\tilde{G}_{\mathcal{R}^{a}_{r}}$, we can conclude:

\begin{enumerate}
\item The family $\left\{ \tilde{G}_{\mathcal{R}} \right\}$ is a commuting set
of operators.
\item They are self-adjoint with respect to the SU(N) Haar measure.
\item Their eigenfunctions are Schur polynomials $\chi_{\cR}(a_j)$.
\end{enumerate}

Lastly it is worth mentioning that, since the q-difference operators are defined for more general root systems, a refined version of \eqref{surfopvsqdiff} might be valid for theories with more general gauge groups.

\bibliographystyle{JHEP}
\bibliography{index}

\end{document}